\documentstyle[prl,multicol,aps,epsfig]{revtex}
\newcommand \beq{\begin{eqnarray}}
\newcommand \eeq{\end{eqnarray}}
\newcommand \bea{\begin{eqnarray}}
\newcommand \eea{\end{eqnarray}}
\def\simge{\mathrel{%
       \rlap{\raise 0.511ex \hbox{$>$}}{\lower 0.511ex \hbox{$\sim$}}}}
\def\simle{\mathrel{
       \rlap{\raise 0.511ex \hbox{$<$}}{\lower 0.511ex \hbox{$\sim$}}}}

\begin{document}

\title{Condensate density and superfluid mass density of a dilute
    Bose gas near the condensation transition}
\author{Markus Holzmann\footnote{Current address:
Laboratoire de Physique Th\'eorique des Liquides, UMR 7600 of CNRS,
Universit\'e P. et M. Curie, boite 121, 4 Place Jussieu, F-75252 Paris}
 and Gordon Baym}
\address{University of Illinois at Urbana-Champaign,
   1110 W. Green St., Urbana, Il 61801, USA}
\maketitle

\begin{abstract}

    We derive, through analysis of the structure of diagrammatic perturbation
theory, the scaling behavior of the condensate and superfluid mass density of
a dilute Bose gas just below the condensation transition.  Sufficiently below
the critical temperature, $T_c$, the system is governed by the mean field
(Bogoliubov) description of the particle excitations.  Close to $T_c$,
however, mean field breaks down and the system undergoes a second order phase
transition, rather than the first order transition predicted in Bogoliubov
theory.  Both condensation and superfluidity occur at the same critical
temperature, $T_c$ and have similar scaling functions below $T_c$, but
different finite size scaling at $T_c$ to leading order in the system size.
Through a simple self-consistent two loop calculation we derive the critical
exponent for the condensate fraction, $2\beta\simeq 0.66$.

\end{abstract}
\begin{multicols}{2}

    The behavior of a dilute interacting homogeneous Bose gas at the
condensation temperature, $T_c$, and in its neighborhood, is a delicate
critical problem [1-3, and references therein].  Even though in a dilute gas,
where $na^3 \ll 1$, with $n$ the number density, and $a$ the s-wave
particle-particle scattering length, perturbation theory fails close to $T_c$,
and fluctuations cannot be neglected; e.g., although the shift in $T_c$ from
the free gas transition temperature, $T_c^0$, is linear in $a$, the first
order perturbation result vanishes, while all higher order terms are
divergent.

    We calculate here the dependence on $a$ of the condensate density, $n_0$,
and the superfluid mass density, $\rho_s$, of a dilute Bose gas, for
temperatures $T$ just below $T_c$, where $t\equiv(T_c-T)/T_c$ is of order
$a/\lambda$.  Here, $\lambda=(2\pi/mT)^{1/2}$ is the thermal wavelength in
units with $\hbar=1$, and $m$ is the particle mass.  As long established, the
Bogoliubov (mean field) approximation fails close to $T_c$; it leads to a
first order phase transition (e.g.  \cite{BG}), vs. the second order
transition expected for the universality class of the Bose gas.  A similar
phenomenon occurs in relativistic $\phi^4$ theory \cite{BG}.  We derive the
general scaling structure of both $n_0$ and $\rho_s$ in the critical region,
as functions of $t$ and $a/\lambda$, which connects the mean field solution to
the critical behavior of a continuous phase transition.  The scaling functions
for $n_0$ and $\rho_s$ are similar, and imply that in the (dilute) interacting
Bose gas the phenomena of condensation and superfluidity occur at precisely
the same temperature.  A further consequence is that in the very dilute limit,
$a \to 0$, where the shift of the critical temperature is linear, $T_c-T_c^0
\sim a$, \cite{NLO,others}, the condensate and superfluid mass densities at
the ideal gas critical temperature, $T_c^0$, both vary linearly with $a$.  We
also calculate the leading order finite-size corrections to $n_0$ and $\rho_s$
at $T_c$, which provides insight into the difference of the numerical results
of Refs~\cite{others} and \cite{linres} for the shift of the critical
temperature.

    Our approach is to study the particle densities via diagrammatic
perturbation theory, working to the lowest needed order in $a$ and $n_0$.  To
deduce the behavior of the superfluid mass density we employ Josephson's
relation between $\rho_s$, $n_0$, and the long wavelength limit of the single
particle Green's function \cite{josephson,standrews,griffin}.

    The particle density, $n$, in the condensed phase is a function of $a$,
$n_0$, and $T$, and has the form, $n(a,n_0,T)=n_0 + \tilde{n}(a,n_0,T)$, where
$\tilde{n}(a,n_0,T)$ is the density of non-condensed particles (with momentum
$k \ne0$).  At the transition temperature, $\tilde{n}(a,0,T_c)=n$; thus
writing $\Delta \tilde{n} \equiv \tilde{n}(a,n_0,T)-\tilde{n}(a,0,T)$, we have
$n_0 + \Delta \tilde{n} = \tilde{n}(a,0,T_c) - \tilde{n}(a,0,T)$.  Since the
difference of $T_c$ from the ideal gas transition temperature, $T_c^0$, is of
order $a$ \cite{club}, we may, to lowest order in $a$, for $t$ of order
$a/\lambda$, replace the difference on the right side by $\tilde{n}(0,0,T_c) -
\tilde{n}(0,0,T) = n\left((T_c/T)^{3/2}-1)\right) \simeq \frac{3}{2}n \, t$.
Up to corrections of order $a^2$, we have then
\begin{equation}
   n_0 + \Delta \tilde{n} =\frac{3}{2}n \, t.
\label{basic}
\end{equation}
Equation (\ref{basic}) implicitly determines the condensate fraction,
$n_0(a,t)$, as a function of $a$ and $t$ in the critical region, $t \simle
a/\lambda$ or $n_0/n \simle a/\lambda$.

    It is simplest to calculate $\tilde{n}(a,n_0,T)$ in terms of the matrix
Green's function, ${\cal{G}}(rt,r't')= -i \left(\langle
T\left(\Psi(rt)\Psi^\dagger(r't')\right)\rangle - \langle
\Psi^\dagger(r't')\rangle \langle \Psi(rt)\rangle\right) $, where the two
component field operator is $\Psi(rt)=\left(\psi(rt),\psi^\dagger(rt)
\right)$.  The Fourier components of ${\cal{G}}^{-1}$ have the form,
\begin{equation}
  {\cal{G}}^{-1}(k,z_n) =
   \pmatrix{
 z_n +\mu-\varepsilon_k -\Sigma_{11} & -\Sigma_{12}\cr
  -\Sigma_{21}& -z_n + \mu-\varepsilon_k -\Sigma_{22} \cr }
\label{gamma}
\end{equation}
where the $z_n\equiv 2\pi n T$ are Matsubara frequencies $(n=0,\pm 1,\pm 2
\dots)$,
$\varepsilon_k=k^2/2m$,
and the $\Sigma_{ij}(k,z_n)$ are the corresponding self-energies.  The
chemical potential, $\mu$, depends here on $n_0$.

    The non-condensate density, $\tilde{n}$, is then found from
\begin{equation}
  \tilde{n}(a,n_0,T) = -T
  \sum_n \int \frac{d^3 k}{(2\pi)^3} G_{11}(k,z_n),
\label{ntilde}
\end{equation}
with
\begin{eqnarray}
  G_{11}&&(k,z) = \nonumber\\
   &&\frac{z -\mu +\varepsilon_k +\Sigma_{22}}{
  (z+\mu-\varepsilon_k-\Sigma_{11})
  (z-\mu+\varepsilon_k+\Sigma_{22})
  +\Sigma_{12}\Sigma_{21}}.
  \label{g11}
\end{eqnarray}
Quite generally, $\Delta \tilde{n}$, to leading order in $a$ and $n_0$, is
given by the $z_n=0$ contribution only:
\begin{equation}
 \Delta \tilde{n}
  =-T \int \frac{d^3 k}{(2\pi)^3} \left[ G_{11}(k,0) - G(k,0) \right],
\label{deltan}
\end{equation}
where $G(k,0)\equiv\lim_{n_0 \to 0} G_{11}(k,0)$; integrated over $k$,
$G(k,0)$ determines the leading order shift of the critical temperature due to
interactions \cite{club}.
The Hugenholtz-Pines relation \cite{HP},
\beq
  \mu = \Sigma_{11}(0,0)-\Sigma_{12}(0,0),
  \label{mu}
\eeq
specifies $\mu$ as a function of $n_0$.  In the zero frequency sector,
$\Sigma_{11}(k,0)=\Sigma_{22}(k,0)$ and $\Sigma_{12}(k,0)=\Sigma_{21}(k,0)$,
so that $ \lim_{k\to0} \left[\left(\mu-\Sigma_{11}(k,0)\right)
\left(\mu-\Sigma_{22}(k,0)\right)\right.$
$\left.-\Sigma_{12}(k,0)\Sigma_{12}(k,0)\right] =0$, and thus the excitation
spectrum is gapless.  In the following we drop the Matsubara frequency index,
always referring to the zero frequency components.

    The lowest order mean field self-energies, $\Sigma_{11}=\Sigma_{11}^{mf}=2
g (n_0 + \tilde{n})$, $\Sigma_{12}=\Sigma_{12}^{mf}=g n_0$, and $\mu = g(n_0+2
\tilde{n})$, where $g=4\pi a/m$, lead to the usual gapless Bogoliubov
excitation spectrum.  The mean field contribution to $\Delta \tilde{n}$, from
Eq.~(\ref{deltan}), is
\begin{equation}
  \Delta \tilde{n}_{mf}=
  -\frac{2}{\pi \lambda^2} \int dk
  \frac{\Sigma^{mf}_{12}}{\varepsilon_k +2 \Sigma^{mf}_{12}}
   = - \frac{2\pi^{1/2}}{\lambda^3} \left( n_0 \lambda^2 a \right)^{1/2}.
 \label{mfdeltan}
\end{equation}
Since the contribution of this term in Eq.~(\ref{basic}) is $\propto
-n_0^{1/2}$ we find two possible solutions of (\ref{basic}) for $n_0$ at the
mean field critical temperature, $t_{mf}=0$, namely $n_0=0$ and $n_0=4 \pi
a/\lambda^4$; intermediate values are not possible for $t>0$.  Thus
Bogoliubov theory predicts a first order phase transition with a jump of the
condensate density from 0 to $n_0=4\pi a/\lambda^4$ \cite{BG}.  However, as we
discuss below, mean field can be valid only outside the critical region
(where $a/\lambda \ll |t| \ll 1$, and thus from Eq.~(\ref{basic}), $a \ll n_0
\lambda^4$), where it implies that $n_0 \propto t$.

    To go beyond mean field we analyze the structure of the self-energies by
expanding $\Sigma_{11}-\Sigma_{11}^{mf}$ and $\Sigma_{12}-\Sigma_{12}^{mf}$ in
a series in $a$ and the mean field Green's functions $G_{11}^{mf}$ and
$G_{12}^{mf}$, given by Eq.~(\ref{gamma}) with the $\Sigma_{ij}$ replaced by
$\Sigma_{ij}^{mf}$.  We eliminate $\mu$ in ${\cal{G}}$ in favor of $n_0$ using
Eq.~(\ref{mu}).  However, rather than using the gapless spectrum directly in a
perturbative expansion, we write the propagators in terms of the mean field
correlation length, $\zeta$, as in \cite{club}, defined by
\begin{eqnarray}
 \mu-(\Sigma_{11}^{mf}-\Sigma_{12}^{mf})
  &=&(\Sigma_{11}(0)-\Sigma_{12}(0))-
   (\Sigma_{11}^{mf}-\Sigma_{12}^{mf}) \nonumber\\
   \equiv -1/2m \zeta^2
\label{HPself}
\end{eqnarray}

    Since the propagators remain formally infrared convergent we can derive
the scaling structure of the self-energies by power-counting arguments.  As
above the transition, the ultraviolet part, when we neglect non-zero Matsubara
contributions, has only a harmless logarithmic divergence which can be removed
by renormalization \cite{club}.  The expansion of the self-energies beyond
mean field starts at order $a^2$; furthermore $\Sigma_{12}$ is formally at
least of order $n_0$.  Diagrams of order $a^\kappa$ with $\kappa \ge 3$ in the
formal expansion contain vertices with two Green's functions entering; similar
to the structure at $T_c$, they involve the dimensionless combinations $a
\zeta/\lambda^2$ and $n_0 \lambda^2 \zeta$.  The latter part originates from
the dependence of ${\cal G}^{mf}$ on $2 m\Sigma_{12}^{mf}\sim an_0$.  Any
diagram with an explicit power, $p$, of $n_0$ can be generated from a
corresponding diagram of power $p-1$ in which a line is replaced by $\sqrt
n_0$ at each of its ends.  Thus each power of $n_0$ involves one fewer
three-momentum loop to be integrated over, replacing a structure of the form
$2m T\int d^3k/(k^2+\zeta^{-2})\sim \zeta^{-1} \lambda^{-2}$, in a loop
integral of order $(a\zeta/\lambda^2)^2$, by $n_0$.  The explicit $n_0$
dependences therefore enter in the combination $(a\zeta/\lambda^2)^2(n_0
\zeta\lambda^2) = a^2n_0\zeta^3/\lambda^2$.  Then with all momenta $k$ scaled
by $1/\zeta$, we find the following scaling structure for the self-energies,
\begin{eqnarray}
   \Sigma_{ij}(k)-\Sigma_{ij}^{mf}(0)
  & =  &  T   \frac{a^2}{\lambda^2} \sigma_{ij}
  \left(k \zeta, \frac{a \zeta}{\lambda^2}, n_0 \lambda^2\zeta \right),
 \label{sigma}
\end{eqnarray}
where the $\sigma_{ij}$ are dimensionless functions of dimensionless
variables. In particular, for vanishing $k$,
\begin{eqnarray}
  ( \Sigma_{11}(0)-\Sigma_{12}(0))
   -( \Sigma_{11}^{mf}(0)-\Sigma_{12}^{mf}(0)) \nonumber\\
   =T \frac{a^2}{\lambda^2} s\left( \frac{a\zeta}{\lambda^2},n_0
 \lambda^2 \zeta
\right),
\label{self2}
\end{eqnarray}
where $s$ is a dimensionless function.

   Equations (\ref{HPself}) and (\ref{self2}) imply that
\begin{equation}
   \zeta = \frac{\lambda^2}{a} h(n_0\lambda^4/a),
  \label{zeta}
\end{equation}
where $h$ is a dimensionless function.  Then using Eqs.~(\ref{sigma}) and
(\ref{zeta}) in (\ref{deltan}), we see that $\Delta \tilde{n}$ has the scaling
structure $\Delta \tilde{n} = (a/\lambda^4) \tilde{f}(n_0 \lambda^4/a)$.  It
immediately follows that close to $T_c$, where $n_0 \sim a/\lambda^4$, the
dimensionless function $\tilde{f}$ cannot be determined by a perturbation
expansion in $n_0 \lambda^3$ or $a/\lambda$; therefore the predictions of mean
field theory fail in this region.  Finally from Eq.~(\ref{basic}), using $n
\sim 1/\lambda^3$ to lowest order, we derive the basic scaling result in
the critical region,
\begin{equation}
  \frac{n_0}{n} = \frac{a}{\lambda} f\left( \frac{t \lambda}{a} \right),
  \quad t \simle \frac{a}{\lambda},
  \label{scaling}
\end{equation}
where $f$ is a dimensionless function.  In the mean field limit,
$x\to\infty$, we must have $f(x\to \infty)\sim x$, whereas the theory of
critical phenomena implies a power-law behavior in the opposite limit, $f(x
\to 0)\sim x^{2 \beta}$, where $\beta$ is the critical index for the order
parameter, $\langle \psi \rangle = \sqrt n_0$.

   We see that for constant $t\lambda/a$, $n_0$ varies linearly with $a$.  As
$a \to 0$, $T_c$ varies linearly in $a$ \cite{club}; thus at the ideal gas
transition temperature $t = t_0\equiv (T_c-T_c^0)/T_c \sim a$, for $a \to 0$,
the condensate fraction varies linearly with $a$.  This result is consistent
with Leggett's weak variational bound that at $t_0$, $n_0$ is bounded above by
terms of order $a^{1/3}$ \cite{tony}.

   We now calculate the scaling function $f(x)$ explicitly within a simple
model beyond the Bogoliubov approximation.  Introducing $U(k) \equiv
2m(\Sigma_{11}(k) - \Sigma_{11}(0))$, we have $\varepsilon_k +
\Sigma_{11}(k) -\mu= (k^2+U(k)+2m\Sigma_{12})/2m$.  Taking for
$\Sigma_{12}(k)$
only the first order diagram in $n_0$, $\Sigma_{12} \equiv \Sigma_{12}^{mf}=g
n_0$, we may write Eq.~(\ref{deltan}) as
\begin{eqnarray}
 \Delta \tilde{n}
  & \simeq & -\frac{4 mgn_0}{\pi \lambda^2} \nonumber \\
  &&\int_0^{\infty} dk
  \frac{k^2}{\left(k^2+U(k)\right)\left(k^2+U(k)+4mgn_0\right)}.
  \label{Udeltan}
\end{eqnarray}
Were we to neglect $U(k)$, we would derive the mean field result
(\ref{mfdeltan}); rather, we determine $U(k)$ from a self-consistent two-loop
calculation, as in \cite{club},
\begin{eqnarray}
   U(k) = -4m g^2 T^2&& \int \frac{d^3 q}{(2\pi)^3}
 \int \frac{d^3 p}{(2\pi)^3} \nonumber\\ &&G(p+q) G(p) \left[ G(k-q) - G(q)
\right],
\end{eqnarray}
where $G^{-1}(k)=\mu - \varepsilon_k- U(k)/2m$ is the inverse of the
($z_n=0$) Green's function at the transition temperature.  A free particle
spectrum in $G$ would leads to a logarithmic infrared divergence.  As in the
calculation of the critical temperature, self-consistency of $U(k)$ at this
level implies that it has the approximate structure, $U(k)= k_c^{1/2}
k^{3/2}$, for non-zero $k \simle k_c$, with $k_c =32(2\pi/15)^{1/2}
a/\lambda^2 \approx 20.7 a/\lambda^2$ \cite{club}.  A change of the power-law
behavior of the free propagator to $G(k)\sim -k^{-(2-\eta)}$, for $k \to 0$,
with $\eta > 0$, is only possible precisely at $T_c$, where the correlation
length diverges.  As long as $n_0\ne 0$, the Bogoliubov operator inequality
\cite{standrews}, $-G_{11}(k,0) \ge m n_0/n k^2$, does not permit $\eta > 0$.
Therefore the $z_n=0$ spectrum remains quadratic, $\eta =0$, as $k \to 0$,
everywhere below and above the critical temperature.

    However, for $n_0 \lambda^4 \ll a$, the details of the spectrum at small
$k$ do not enter, and we may approximate the self-energy as:
\begin{eqnarray}
  U(k) =
 \left\{
 \begin{array} {l@{\quad : \quad}l}
  k_c^{1/2} k^{3/2} &  k \ll k_c \\
  k_c^2 & k \gg k_c,
\end{array}
\right.
\end{eqnarray}
which in Eq.~(\ref{Udeltan}) leads to
\begin{eqnarray}
 \Delta \tilde{n} & \simeq &
 \left\{
\begin{array}{l@{\;:\;}l}
  (32 n_0 a/3 k_c \lambda^2)
 \ln (16 \pi n_0 a/k_c^2) & n_0\lambda^4 \ll a \\
-(2\pi^{1/2}/\lambda^3)\left( n_0 \lambda^2 a \right)^{1/2}
 & n_0 \lambda^4 \gg a.
 \end{array} \right.
\label{model}
\end{eqnarray}
The second line is the mean field result (\ref{mfdeltan}).  Taken
literally, this model would again predict a first order phase transition;
however the logarithmic term indicates a change in the power law behavior
close to the critical point.  To determine this relation we invert
Eq.~(\ref{scaling}), using $n\sim \lambda^{-3}$ to note that $tn/n_0$ must be
a dimensionless function of the variable $\lambda^4n_0/a$.  In the limit $n_0
\lambda^4 \ll a$, we may write, using the upper result in Eq.~(\ref{model}),
  \begin{equation}
   n_0+ \Delta \tilde{n} =\frac32 nt \simeq n_0 \left( 1
             + \frac{32 a}{3 k_c \lambda^2} \ln
  \left(\frac{16 \pi n_0 a}{k_c^2} \right) \right),
\end{equation}
which is the first term in an expansion of the scaling form
\begin{equation}
   nt \sim n_0\left(\frac{\lambda^4 n_0}{a}\right)^{32a/3k_c\lambda^2},
\end{equation}
in the formal limit $a/k_c\lambda^2 \to 0$, consistent with our
approximation of $U(k)$.  Inverting, we derive $n_0 \sim t^{2\beta}$, with
$2\beta = {1/(1+(5/6\pi)^{1/2})}\simeq 0.66$, in excellent agreement with the
value $2\beta \simeq 2/3$ expected for this universality class.  In the other
limit, $n_0 \lambda^4 \gg a$ this model calculation simply approaches the mean
field result, $n_0 \sim t$.

   Below $T_c$ the condensed system is superfluid, with a superfluid mass
density, $\rho_s$, related to $n_0$ by Josephson's sum rule
\cite{josephson,standrews,griffin},
\begin{eqnarray}
 \rho_s = -\lim_{k \to 0} \frac{ n_0 m^2 }{k^2 G_{11}(k,0)}.
\label{josephson}
\end{eqnarray}
Using the explicit form (\ref{g11}) for $G_{11}(k,0)$, we have
\begin{equation}
  \rho_s = n_0m +2n_0m^2\frac{\partial}{\partial k_z^2}
  (\Sigma_{11}(k)-\Sigma_{12}(k))\Big|_{k=0},
 \label{rhos}
\end{equation}
for $T \le T_c$.  This result implies that precisely at $T_c$, the
superfluid fraction vanishes with the condensate fraction.  Above $T_c$, the
superfluid density vanishes, as can be directly derived by calculating the
transverse current-current correlation function \cite{transverse}.

   Further, we immediately see from Eqs.~(\ref{rhos}) and (\ref{sigma}) that
the superfluid fraction in the neighborhood of the ideal gas transition has
the same scaling behavior as we found for the condensate density:
\begin{equation}
  \frac{\rho_s}{mn}  = \frac{a}{\lambda} f_{\rho}\left( \frac{t \lambda}{a}
\right),
  \quad t \simle \frac{a}{\lambda};
  \label{rhoscaling}
\end{equation}
however the scaling function $f_{\rho}$ is in general different from
the scaling function $f$.  Reference \cite{rasolt} obtained the scaling
function for $\rho_s$ in the dilute limit to order $\epsilon=4-d$, where $d=3$
is the spatial dimension.  In the mean field limit, $t\gg a/\lambda$, the
 lowest order self-energies are independent of $k^2$, so that from
Eq.~(\ref{rhos}) the superfluid mass density coincides with $n_0$ to order
$a$.  Thus $f_{\rho}( x\to\infty)\sim x$.  In the critical region, however,
$f_{\rho}(x \to 0)\sim x^{\gamma}$, where $\gamma$ is the critical
index for the superfluid mass density.  Josephson's scaling relation gives
$\gamma = 2 \beta-\eta \nu \simeq 2/3$ for the critical index of the
superfluid mass density, where $\nu\simeq 2/3$ is the critical exponent of the
correlation length \cite{josephson}.  Since our model calculation above does
not include the correct $k \to 0$ limit, to which $\rho_s$ (but not $n_0$) is
sensitive, it is therefore not suitable for calculating $\rho_s$ reliably.

   Let us turn to understanding the behavior of $n_0$ and $\rho_s$ in large
but finite systems, of linear scale $L$.  The condensate density,
$n_0^L=n-\tilde{n}^L$ at $T_c$ is non-zero for finite $L$, and is found to
leading order from
\begin{eqnarray}
  n &=& n_0^{\infty}-T \int_0^{\infty}
\frac{d^3k}{(2\pi)^3}G(k,\zeta)
\nonumber\\   &=&
  n_0^{L}-T \int_{2\pi/L}^{\infty} \frac{d^3k}{(2\pi)^3}G(k,\zeta),
\end{eqnarray}
where $G$ is the infinite size Green's function.  Since at $T_c$, $G(k\to
0)= -2mC \zeta^\eta/k^{2-\eta}$, where $C$ is constant, and $n_0^\infty=0$, we
find,
\begin{equation}
   n_0^L = \frac{4C}{(1+\eta)\lambda^2 L}
   \left(\frac{2 \pi \zeta}{L}\right)^{\eta},
  \label{n0L}
\end{equation}
to leading order, neglecting a numerical factor dependent on the
particular geometry of the finite system.  Josephson's relation should still
hold inside the critical region of finite size systems, with the limit of zero
wavevector replaced by $k\to 2\pi/L$.  With this relation we have
\begin{equation}
   N^{1/3}\frac{\rho_s}{m n}= \frac{2}{(1+\eta)\lambda^2 n^{2/3}}=
   \frac{T_c}{T_c^0}\frac{2 }{(1+\eta)(\zeta(3/2)^{2/3}},
  \label{rhosL}
\end{equation}
where the total particle number is given by $N=nL^3$.  Note that both
Eqs.~(\ref{n0L}) and (\ref{rhosL}) are valid independent of the diluteness of
the gas.  Equation (\ref{rhosL}) agrees well with the numerical values of
Ref.~\cite{peter}.  Since the limit as $a \to 0$ of $\eta$ is non-zero for an
interacting Bose gas, the formal $a\to 0$ limit of Eq.~(\ref{rhosL}) does not,
however, agree with the ideal gas value, given by the same formula but with
$\eta=0$.  Therefore, the procedure of Ref.~\cite{linres}, to expand the
finite-size scaling results directly around the ideal gas limit, is not
completely justified; that there was a difficulty in this method was already
suggested by the disagreement of the calculated value of $T_c-T_c^0$ with
later lattice calculations \cite{others} which do not rely on this assumption.
Nevertheless, use of finite-size scaling raises new strategies for explicit
calculations \cite{erich}.

   We would like to thank the Aspen Center for Physics where this work was
initially conceived and finally completed.  MH acknowledges seminal
discussions with David Ceperley on the finite size corrections.  This research
was supported in part by the NASA Microgravity Research Division, Fundamental
Physics Program, and by NSF Grants PHY98-00978 and PHY00-98353.

\end{multicols}

\begin{thebibliography}{99}

    \bibitem{club} G. Baym, J.-P.  Blaizot, M. Holzmann, F. Lalo\"e, and D.
Vautherin, Phys.  Rev.  Lett.  {\bf 83}, 1703 (1999); Euro.  J. Phys.  B24,
104 (2001).

   \bibitem{gordon2} G. Baym, J.-P.  Blaizot, and J. Zinn-Justin, Europhys.
Lett.  {\bf 49}, 150 (2000);  P. Arnold and B. Tom\'a{\^s}ik,
Phys.  Rev.  A{\bf 62}, 063604 (2000).

   \bibitem{others} P. Arnold and G. Moore, Phys.  Rev.  Lett.  {\bf 87},
120401 (2001), V.A.  Kashurnikov, N.~V.  Prokof'ev, and B.~V.  Svistunov,
Phys.  Rev.  Lett.  {\bf 87}, 120402 (2001)

   \bibitem{BG} G. Baym and G. Grinstein, Phys.\ Rev.\ {\bf D15}, 2897 (1977).

   \bibitem{NLO} M. Holzmann, G. Baym, J.-P.  Blaizot, and F. Lalo\"e, Phys.
Rev.  Lett.  {\bf 87}, 120403 (2001); P. Arnold, G. Moore, and 
B. Tom{\'a}{\^s}ik,
Phys. Rev. {\bf A65}, 013606 (2001). 

    \bibitem{linres} M. Holzmann and W. Krauth, Phys.  Rev.  Lett.  {\bf 83},
2687 (1999).

    \bibitem{josephson} B.D.~Josephson, Phys.~Lett.  {\bf 21}, 608 (1966).

    \bibitem{standrews} G. Baym, in Mathematical Methods in Solid State and
Superfluid Theory, ed. by R.C.  Clark and G.H.  Derrick (Oliver and Boyd,
Edinburgh, 1969), p 121.

    \bibitem{griffin} E. Talbot and A. Griffin, Ann.  Phys.  (New York) {\bf
151}, 71 (1983); Phys.  Rev.  {\bf B 29}, 3952 (1984).

    \bibitem{HP} N.M.  Hugenholtz and D. Pines, Phys.  Rev.  {\bf 116}, 489
(1959).

    \bibitem{tony} A.J.~Leggett, (2002).  New J. Physics {\bf 3}, 23 (2001).

    \bibitem{transverse} The vanishing of $\rho_s$ above and at $T_c$ follows
directly from the definition of the transverse current-current correlation
function.  Since $G^{-1}(k\to 0) \sim k^2$ below $T_c$, Eq.~(\ref{josephson})
implies that both Bose condensation and superfluidity set in at the same
critical temperature, $T_c$, for any interacting Bose fluid, not only the
dilute gas.  We discuss more fully, in a future publication [M.  Holzmann and
G. Baym, to be published], the relation between $\rho_s$, $n_0$ and $G(k)$, as
well as derive the Josephson relation from diagrammatic perturbation theory
and discuss its relation to transverse current-current correlations.

    \bibitem{rasolt} M. Rasolt, M.J. Stephen, M.E. Fisher, and P.B.  Weichman,
Phys.  Rev.  Lett.  {\bf 53}, 798 (1984).

    \bibitem{peter} P. Gr\"uter, D.M. Ceperley, and F. Lalo\"{e}, Phys.  Rev.
Lett.  {\bf 79}, 3549 (1997).

    \bibitem{erich} E. Mueller, G. Baym, and M. Holzmann, J. Phys.  B:  At.
Mol. Opt.  Phys.  {\bf 34}, 4561 (2001)

\end{thebibliography}
\end{document}